\documentclass[aps, prb, numerical, reprint]{revtex4-1}
\usepackage{graphicx}
\usepackage{epstopdf} 
\usepackage{bm}
\usepackage{dcolumn}
\usepackage{amssymb}
\usepackage{amsmath}

\begin{document}

	\title{Origin of the resistance-area product dependence of spin transfer torque switching in perpendicular magnetic random access memory cells}
	
	\author{G.\ Mihajlovi\'{c}}
	\email{goran.mihajlovic@wdc.com}
	\affiliation{Western Digital Research Center, Western Digital Corporation, San Jose, CA 95119\\}
	
	\author{N.\ Smith}
	\affiliation{Western Digital Research Center, Western Digital Corporation, San Jose, CA 95119\\}
	
	\author{T.\ Santos}
	\affiliation{Western Digital Research Center, Western Digital Corporation, San Jose, CA 95119\\}
	
	\author{J.\ Li}
	\affiliation{Western Digital Research Center, Western Digital Corporation, San Jose, CA 95119\\}
	
	\author{M.\ Tran}
	\affiliation{Western Digital Research Center, Western Digital Corporation, San Jose, CA 95119\\}
	
	\author{M.\ Carey}
	\affiliation{Western Digital Research Center, Western Digital Corporation, San Jose, CA 95119\\}
	
	\author{B.\ D.\ Terris}
	\affiliation{Western Digital Research Center, Western Digital Corporation, San Jose, CA 95119\\}
	
	\author{J.\ A.\ Katine}
	\email{jordan.katine@wdc.com}
	\affiliation{Western Digital Research Center, Western Digital Corporation, San Jose, CA 95119\\}
	
	\date{\today}
	
	
\begin{abstract}
		
We report on an experimental study of current induced switching in perpendicular magnetic random access memory (MRAM) cells with variable resistance-area products ($RAs$). Our results show that in addition  to spin transfer torque (STT), current induced self-heating and voltage controlled magnetic anisotropy also contribute to switching and can explain  the $RA$ dependencies of switching current density and STT efficiency. Our findings suggest that thermal optimization of perpendicular MRAM cells can result in significant reduction of switching currents.
		
\end{abstract}
	
\maketitle
	
\section{Introduction}

As information technology enters a new era \cite{TheisandWong_CSE2017}, with Internet of Things expected  to connect over 30 billion devices generating vast amount of data that need to be processed and stored \cite{Marjani_IEEEAccess2017}, there is a rapidly growing demand for faster, denser and more power-efficient non-volatile memories \cite{Meena_NanoResLett2014} that could be organized in alternative hierarchies offering better system performance and greater functionality \cite{Wong&Salahuddin_NatNano2015}, all at preferably lower cost. Spin transfer torque magnetoresistive random access memory (STT MRAM) \cite{Apalkov_ProcIEEE2016, Kent&Worledge_NatNano2015} is uniquely positioned to address this challenge as it is the only emerging memory that could combine the high speed and endurance of SRAM, high density of DRAM and the non-volatility of Flash \cite{Khvalkovskiy_JPD2013}. The heart of the MRAM cell is the magnetic tunnel junction (MTJ), that provides the write, read and bit storing functionality, essentially using two magnetic layers, reference layer (RL) and the free layer (FL), separated by a magnesium oxide (MgO) tunnel barrier \cite{Apalkov_ProcIEEE2016, Khvalkovskiy_JPD2013}. The two bit storage states are the parallel (P) and antiparallel (AP) magnetization orientations of the FL relative to the RL, distinguished by different resistance-area products ($RA$) of the MTJ: $(RA)_P \equiv RA$ for the P state, and $(RA)_{AP} = (1 +TMR)RA$ for AP state, with $TMR$ being the tunneling magnetoresistance ratio.  

For RL and FL with perpendicular magnetic anisotropy (PMA), the STT critical P $ \rightleftharpoons$ AP switching voltage $V_{c0}$ (defined at zero temperature and for infinitely long time) is, in the macrospin approximation,\cite{Sun_PRB2000} expressible in terms of a spin torque field $H_{ST}$ and torquence $\tau$ as

\begin{equation}
\alpha H_k = \pm H_{ST} = \pm \tau V_{c0} /RA,  \tau = \frac{\hbar}{2 e} \frac{\eta}{M_s t },
\label{eq:1}
\end{equation}
where $\alpha$, $M_s$, $t$, and $H_k$ are the damping parameter, saturation magnetization, thickness, and net PMA field of the FL, respectively, and $\eta = \sqrt{TMR(TMR+2)}/(2(TMR+1))$ is a polarization efficiency factor. Apart from  a minor $RA$ dependence of $\eta$, due to  $TMR$ being a weak function of $RA$ (see Table I), the critical current density $J_{c0} \equiv V_{c0}/RA $ is not expected to depend on $RA$. Experimentally, however, an $RA$ dependence has been observed by several groups \cite{Wang_IEEETM2009, Zeng_APL2011, Hu_IEDM2017, Sun_PRB2017} and attributed\cite{Zeng_APL2011, Sun_PRB2017} to an $RA$-dependent spin pumping \cite{Tserkovnyak_PRB2002} contribution to $\alpha$ in Eq.~(1). Here we show that the $RA$ dependence of $J_{c0}$ is influenced by other phenomena, in particular the current-induced self-heating of an MRAM cell which reduces the effective $H_k$ of the FL, and, to a smaller extent, the voltage controlled magnetic anisotropy effect (VCMA) \cite{Maruyama_NatNano2009, Amiri&Wang_SPIN2012}. As the temperature rise of the FL is proportional to the dissipated power density \cite{Papusoi_NJP2008} $RA J^2$, higher $RA$ devices result in lower $J_{c0}$. In addition, as the VCMA effect is proportional to the bias voltage $V_b$ across the MRAM cell, for a given $J$ VCMA effects are stronger with higher $RA$. The combination of heating and VCMA quantitatively explains all of our experimental findings, in particular the much stronger $RA$ dependence of $J_c$ for P to AP switching (P$\rightarrow$AP) compared to AP$\rightarrow$P, and the $RA$ dependence of STT efficiency $E_b/I_{c0}$ obtained from pulse width $t_p$  dependent measurements of switching voltage $V_c$ in the thermally activated (TA) regime \cite{Koch_PRL2004, Li_PRB2004}  ($E_b$ is the energy barrier for magnetization reversal of the FL and $I_{c0} \equiv V_{c0}/R_P$ is the critical switching current). 
		
\begin{table}
\caption{\label{tab:table1} Transport and magnetic properties of free layer films used in this study.}
\begin{ruledtabular}
\begin{tabular}
					{ccccccccc}& $RA$        & $TMR$    & $M_s t$            & $H_k$  & $\alpha$ \\
					& ($\Omega \mu$m$^{2}$)  & (\%)     & (memu/cm$^2$)      & (kOe)  &          \\
					\hline
					& 5                      & 133      & 0.232              & 2.71   & 0.0100   \\
					& 10                     & 147      & 0.227              & 2.72   & 0.0102   \\
					& 15                     & 156      & 0.226              & 2.69   & 0.0100   \\
					& 20                     & 156      & 0.232              & 2.69   & 0.0094   \\
\end{tabular}
\end{ruledtabular}
\end{table}
		
\section{Device fabrication}

The MRAM film stacks used in this study consist of a seed layer, synthetic antiferromagnet RL, MgO tunnel barrier, CoFeB-based FL, MgO cap layer for enhancing $H_k$, and Ru/Ta cap layer. The films were deposited by magnetron sputtering in an Anelva C-7100 system and then annealed at 335$^\circ$C for 1 hour. The MgO layers were rf-sputtered from a MgO target. The $RA$ and $TMR$ values measured on the annealed films by current-in-plane tunneling (CIPT) \cite{Worledge&Trouilloud_APL2003} are shown in Table I. Variation of $RA$ values in the range 5 - 20 $\Omega \mu$m$^{2}$ was achieved by adjusting the sputter time of the MgO barrier, and consequently, the TMR ratio increased from 133 to 156 \%, respectively.  For this range of $RA$ values, $M_s t$ measured by vibrating sample magnetometry,  as well as $H_k$ and $\alpha$ of the FL measured by full film ferromagnetic resonance (FMR) are identical (see Table~I). 
		
\begin{figure}
	\includegraphics [scale = 0.43, bb = 20 10 525 450]{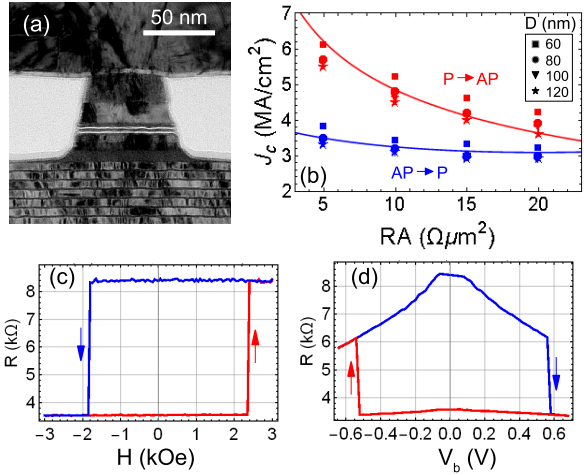}
	\caption{(a) TEM image of an MRAM cell used in this study. (b) Measured $J_c$~vs~$RA$ (symbols) and calculated (lines) using fit parameter values described in text. Each  $J_c$ data point value is median from $>$500 devices. Measured (c) $R$~vs~$H$ and  (d) $R$~vs~$V_b$  of an MRAM cell with $RA10$.}
	\label{fig1:dc results}
\end{figure}
		
MRAM test device cells are fabricated using 193~nm deep UV optical lithography, followed by reactive ion etching a hard mask, ion milling the MRAM film, SiO$_2$ refill and chemical mechanical planarization. Median electrical device diameters $D$, determined by fitting $R_P$~vs~$RA$ for the given optical mask size, are $\sim$120, 100, 80 and 60 nm. A transmission electron microscopy (TEM) image of a representative device is shown in Fig.~1(a). Fig.~1(c) shows $R$ vs. perpendicular external magnetic field $H$ for an MRAM cell with $RA = 10~\Omega \mu$m$^2$ ($RA$10)  and $D \cong 60$~nm measured at constant $V_b =50$~mV, showing $TMR \cong 140$~\%, coercive field $H_c = (H_{SW}^{P \rightarrow AP} - H_{SW}^{AP \rightarrow P})/2 \cong 2$~kOe ($H_{SW}$ is the switching field) and offset field  $H_{offs} = (H_{SW}^{P \rightarrow AP} + H_{SW}^{AP \rightarrow P})/2 \cong 300$~Oe that favors the P state. Fig.~1(d) shows  $R$ vs $V_b$. One can see that P$\rightarrow$AP and AP$\rightarrow$P occur at $V_c^{P \rightarrow AP} = -0.54$~V and $V_c^{AP \rightarrow P} = +0.58$~V, respectively.
		
\section{Results and Analysis}

Fig.~1(b) shows $J_c = I_c/\left( D^2 \pi/4\right) $, determined by ramping $V_b$ with a dwell time of $\sim$10~ms and measuring current $I_c$ just before switching, as a function of $RA$. $J_c$ decreases with increasing $RA$ for both AP$\rightarrow$P and P$\rightarrow$AP. The dependence, however, is much stronger for the latter, with $J_c$ decreasing $\sim$50~\% from $RA$5 to $RA$20, while for AP$\rightarrow$P the decrease is only $\sim15$~\%. Also, $J_c$ at a given $RA$ increases with decreasing $D$. This is contrary to what one would expect in the TA switching regime of these measurements, as smaller devices are more thermally unstable. 
	
The change in $J_c$ with $RA$ cannot be attributed to an $RA$-dependent spin-pumping  \cite{Tserkovnyak_PRB2002} contribution to $\alpha$  as our film FMR measurements show that $\alpha$ is  independent of $RA$ (see Table~I). It also cannot be explained by any dependence of $M_s$ or $H_k$ of the FL on $RA$ as they are also measured to be $RA$-independent (Table~I). In order to understand the origin of these dependencies we performed additional $R$~vs~$H$ measurements as a function of $V_b$.

\begin{figure}
	\includegraphics [scale = 0.48, bb = 0 0 485 475]{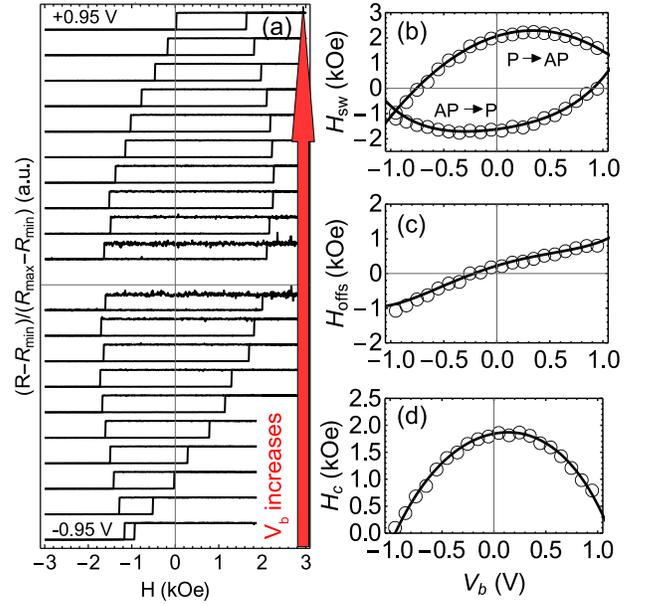}
	\caption{(a) $R$~vs~$H$ for -0.95~V$<V_b<$+0.95~V for a MRAM cell with $RA$20 and  $D = 80$~nm. (b) $H_{SW}$, (c) $H_{offs}$ and (d) $H_c$ vs $V_b$ obtained from the measurements shown in (a) (symbols) and the corresponding dependencies calculated using Eqs.~(2) and (3) (lines) with $H_{c0} = 1.85$~kOe, $H_{RL} =225$~Oe, $\tau/\alpha =18.6 $~kOe$\mu$m$^2$/A, $\epsilon = 0.37$~kOe/V and $\zeta = 37.8$~kOe$\mu$m$^2$/W.}
	\label{fig3:Vb dependence}
\end{figure}
	
Fig.~2(a) shows representative $R$~vs~$H$ data for different $V_b$ from a single cell. $V_b$ is varied from $-0.95$~V (bottom curve) to  $+0.95$~V (top curve) in 0.1~V steps. The obtained $V_b$ dependencies of $H_{SW}$ for P$\rightarrow$AP and AP$\rightarrow$P, $H_{offs}$ and  $H_c$ are shown in Figs~2(b), 2(c) and 2(d), respectively. While the near-linear $V_b$-dependence of $H_{offs}$ shown in Fig.~2(c) is close to expected from STT \cite{Sun_PRB2000}, Fig.~2(d) shows that  $H_c$ exhibits a quadratic component of $V_b$-dependence that strongly suggests self-heating. Indeed, in the macrospin approximation \cite{Sun_PRB2000}, STT alone predicts no dependence of $H_c$ on $V_b$. A more careful inspection of Fig.~2(d) shows that $H_c$ also exhibits a smaller linear component of $V_b$-dependence, which could be due to VCMA. 

\begin{figure}
\includegraphics [scale = .53, bb = 5 5 450 280]{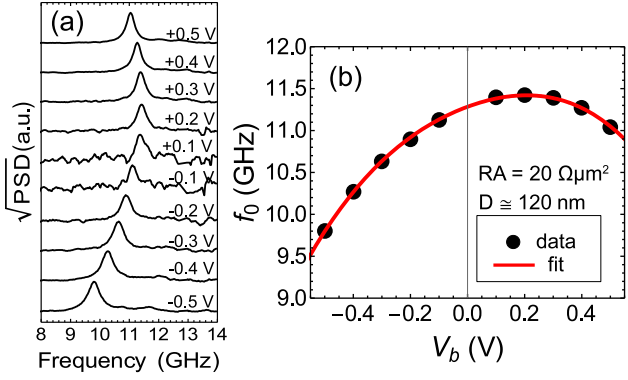}
\caption{(a) Normalized mag-noise root mean square power spectral density measured for different $V_b$. (b) The resonance frequency $f_0$ vs $V_b$ corresponding to the measurements shown in (a). The line is fit to the data using a model that includes VCMA and self-heating contributions, as described in text.}
\label{fig4:FMR}
\end{figure} 
	
Alternatively, the $V_b$-dependence of $H_k$ can be measured more directly (see Fig.~3) from device-level thermally induced FMR (mag-noise) spectra \cite{Smith_PRB2010}. The expected peak resonance frequency $f_0 \approx \gamma \sqrt{\left( \left(H_k\pm H_z\right)^2 + H_{ST}^2\right)  \left( 1-\left(H_y/H_k\right)^2 \right)}$ where $\gamma \cong 3$~GHz/kOe is the gyromagnetic ratio, $H_y$ and $H_z$ are the total in-plane and perpendicular magnetic fields, respectively, and  $H_{ST}$ = $(\alpha H_k)(V_b/V_{c0})$ (see Eq.~1). For the measurements in Fig.~3 (near the AP state), $V_b < V_{c0}$, thus $H_{ST}$ is negligible, $H_z \approx 0$, and $H_y \cong 1$~kOe $\ll f_0/\gamma$ makes only a small correction to $H_k$.  As shown in Fig.~3 for  an $RA$20 cell, $f_0(V_b)$ has both a quadratic and linear (VCMA) contributions, the latter more clearly visible than indicated by   $H_c$ vs $V_b$ shown in Fig.~2(d). One can fit this dependence by expressing $H_k = H_{k0} + \epsilon V_b - \zeta V_b^2/RA'$ where $RA'= RA(1 + TMR|_{V_b = 0})(1-0.5|V_b|)$ is the approximate expression for $V_b$-dependent $RA$ in the AP state (see Fig.~1(d)). The values obtained are $H_{k0} = (3.76 \pm 0.01)$~kOe, $\epsilon = (0.42 \pm 0.01)$~kOe/V and $\zeta = 44.0 \pm 0.5$~kOe$\mu$m$^2$/W. The sign of the VCMA is positive, i.e. it increases $H_k$ for positive $V_b$ (AP$\rightarrow$P polarity). 
	
Having established that VCMA and self-heating are present, $H_{SW}(V_b)$ is explicitly expressed as  
\begin{equation}
H_{SW}^{P \rightarrow AP} = H_{c0} + H_{RL} + \frac{\tau}{\alpha}  \frac{V_b}{RA} + \epsilon V_b - \zeta \frac{V_b^2}{RA},
\label{eq:2}
\end{equation}
	
\begin{equation}
H_{SW}^{AP \rightarrow P} = -H_{c0} + H_{RL} + \frac{\tau}{\alpha}  \frac{V_b}{RA} - \epsilon V_b + \zeta \frac{V_b^2}{RA'},
\label{eq:3}
\end{equation}
where $H_{RL}$ characterizes the perpendicular dipolar stray field from the reference layer and the following terms are from STT, VCMA, and self-heating effects, respectively.
	
\begin{figure}
	\includegraphics [scale = 0.58, bb = 5 0 405 380]{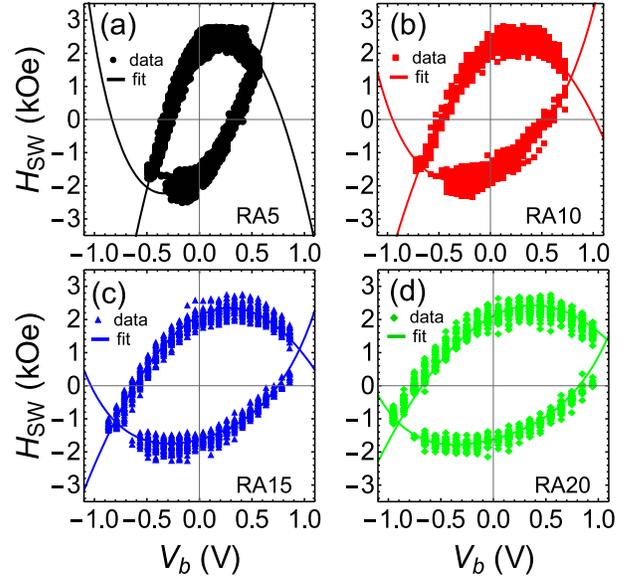}
	\caption{$H_{SW}$ vs $V_b$ for different $RA$. The data on each plot is from all measured devices ($\sim$40 in total) with $D = 60$ - 120 nm. Lines are fits to Eqs.~(2) and (3) with simultaneous fit parameters.}
	\label{fig5:HswVbfit}
\end{figure}  
	
Figs.~4(a)-4(d) show $\textit{simultaneous}$ fits to $H_{SW}^{P \rightarrow AP}$ and $H_{SW}^{AP \rightarrow P}$ vs $V_b$ for $\textit{all}$ $RA$ values explored in this experiment. All data can be fitted with the same set of $RA$-$\textit{independent}$ parameters:  $H_{c0} = (1.86 \pm 0.01)$~kOe, $H_{RL} = (244 \pm 3)$~Oe, $\tau/\alpha = (18.6 \pm 0.1)$~kOe$\mu$m$^2$/A, $\epsilon = (0.42 \pm 0.01)$~kOe/V and $\zeta = (42.8 \pm 0.2)$~kOe$\mu$m$^2$/W.

One can now calculate $V_c$ by solving Eqs.~(2) and (3) for $ V_b$ for which $H_{SW} = 0$. Then  $J_c^{P \rightarrow AP} = V_c^{P \rightarrow AP}/RA$ and $J_c^{AP \rightarrow P} = V_c^{AP \rightarrow P}/RA'(V_c^{AP \rightarrow P})$. The calculated $J_c$ dependencies on $RA$ using the $RA$-independent fit parameters are shown as lines in Fig.~1(b). The agreement is excellent for both $P \rightarrow AP$  and AP$\rightarrow$P. In particular, the model reproduces the much stronger  $RA$ dependence of $J_c$ for  P$\rightarrow$AP.
	
The mild increase of $J_c$  with decreasing cell size shown in Fig.~1(b) is believed to result from more relative cell cooling via three-dimensional heat flow into the surrounding encapsulation material, in addition to weakly increasing $H_{c0}$ with decreasing device size due to reduced demagnetization field near the FL edges \cite{Sun_PRB2013}. The deviation of  $H_{offs}$ from linear dependence on $V_b$ as shown in Fig.~3(c) arises from the differences in the self-heating terms in Eqs.~(2) and (3) for P$\rightarrow$AP and  AP$\rightarrow$P, respectively.

The value of $H_{k0} \cong$ 3.8~kOe extracted from the FMR data of Fig.~3 is a factor of two larger than the value of $H_{c0} \cong $ 1.9 kOe characteristic of the Fig.~4 data. The former is a passive measurement under quiescent macrospin conditions, and should better represent the true device FL PMA compared to the latter, which likely involves a nucleated magnetization reversal process \cite{Shaw_PRB2008} not resembling uniform macrospin rotation. In the macrospin picture (see Eq.~1), $\tau/\alpha$ = $H_{k0}/J_{c0}$ = $\left( \hbar \eta \right) /\left( 2 e \alpha M_s t\right)$. Using Table I, one then estimates $\tau/\alpha \cong$ 65~kOe$\mu$m$^2$/A. This is about 3.5 times larger than the value found from fitting the data in Fig.~4. More than half of this discrepancy may be ascribed to the aforementioned factor of two difference between macrospin $H_{k0}$ and $H_{c0}$ obtained by fitting the same non-macrospin data of Fig.~4 used to fit $\tau/\alpha$.

\begin{figure}
	\includegraphics[scale = 0.48, bb = 10 0 480 380]{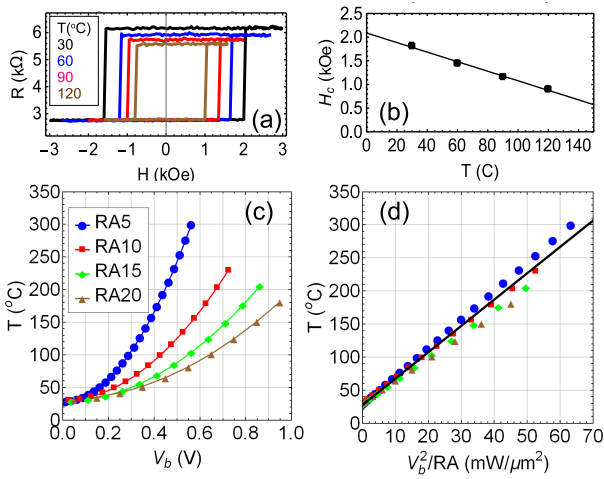}
	\caption{(a) Measured $R$ vs $H$ for different $T$s for an MRAM cell with $RA$20, $D = 100$~nm. (b) $H_c$ vs $T$ for the data shown in (a) (symbols) and linear fit (line) with slope $dH_c/dT \cong 10$~Oe/K. (c) FL $T$ vs $V_b$ determined for the P state for different $RA$s. (d) $T$ vs dissipated power density for different $RA$s. The line is linear fit to the data, i.e. $T = T_0 + R_{th}A V_b^2/RA$ with $T_0 = (28 \pm 2) ^o$C and $R_{th}A = (4.0 \pm 0.1) $~K$\mu$m$^2$/mW. Each data point in (c) and (d)  is the median from $\sim$25~measured cells averaged over $D$ = 60, 80, 100 and 120~nm devices.}
	\label{fig5:Tdependence}
\end{figure}

In order to determine how the cell temperature $T$ depends on $V_b$, we performed $R$ vs $H$  measurements over  $T$ range  $30 - 120 ^o$C. Figs.~5(a) and 5(b) show representative results obtained from single cell. A typical value $dH_c/dT \cong 10$~Oe/K is obtained that is within 10\% of the $dH_k/dT$ found from thermal FMR measurements analogous to those shown in Fig. 3. The measured $dH_c/dT$  factors convert $H_c$ vs $V_b$ data into $T$ vs $V_b$  and $T$ vs $V_b^2/RA$, as is illustrated in the figure and described in the caption.

We also measured $V_c$ vs $t_p$ in the range 10~ns to 5~ms and evaluated $J_{c0}$, thermal stability factor $\Delta = E_b/k_B T$ ($k_B$  is the Boltzmann constant) and  $E_b/I_{c0}$ using the TA model \cite{Li_PRB2004, Koch_PRL2004}. Fig.~6(a) shows an example of the data from a $RA$10 cell, which in  the range $t_p \geq$~5~$\mu$s is fit to the TA model $\ln \left( t_p/\left( \tau_0 \ln2 \right) \right) = \Delta_{eff} = \Delta \left( H_k/H_{k0} \right) \left( T_0/T\right) $ using the following two forms:
\begin{equation}
H_k  = H_{k0} \left(1 \pm \frac{V_c}{V_{c0}} \right), ~T = T_0
\label{eq:4}
\end{equation}
(solid lines) and 
\begin{equation}
H_k  = H_{k0}  \pm \frac{\tau}{\alpha} \frac{V_c}{RA} + \epsilon V_c - \zeta \frac{V_c^2}{RA'}, ~T = T_0 + R_{th}A \frac{V_c^2}{RA'}
\label{eq:5}
\end{equation} 
(dashed lines) where $\tau_0 = 1$~ns is taken to be the inverse attempt frequency, $H_{k0}$ and $T_0$ are $H_k$ and $T$ at $V_b = 0$, $R_{th}A$ is the effective thermal resistance-area product and $(+)$ and $(-)$ sign correspond to P$\rightarrow$AP and AP$\rightarrow$P, respectively. Eq.~(4) is commonly found in the literature \cite{Koch_PRL2004, Li_PRB2004, Sun_PRB2017} where only STT influence is accounted for, while Eq. (5) incorporates the additional $V_b$ dependencies of $H_k$ from both VCMA and self- heating, as well as the explicit $V_b$ dependence of cell $T$, as described earlier via Eqs.~(2), (3) and Figs.~4 and 5. Along with fit parameter $\Delta$  (both forms), Eq. (4) uses the second fit parameter $V_{c0}$. When using Eq.~(5), $H_{k0}$   is the only additional fit parameter, while the values for $\tau/\alpha$, $\epsilon$ and $\zeta$  are those $RA$-$\textit{independent}$ parameter values determined from the data of Fig.~(4), $T_0 = 303$~K and  $R_{th}A \cong 4$~K$\mu$m$^2$/mW is determined from data in Fig.~5. For Fig.~6(c), $J_{c0} = V_{c0}/RA$ for Eq.~(4) case and $J_{c0} = \left( \tau/\alpha\right)^{-1}/H_{k0}$ for Eq.~(5) case. Note that, in both cases, AP$\rightarrow$P and P$\rightarrow$AP branches are fit separately and $V_{c0}$ and $\Delta$ are determined as their average. One can see in Fig.~6(a) that both models fit the data well (the solid and dashed lines are indistinguishable).

\begin{figure}
	\includegraphics[scale = 0.53, bb = 0 0 410 485]{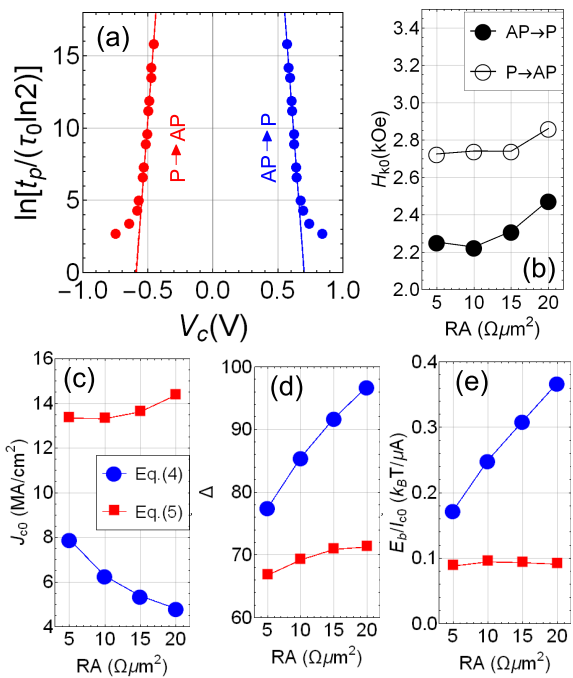}
	\caption{(a) $V_c$ vs $\ln\left( t_p/\left( \tau_0 \ln2 \right) \right) $ values (symbols) measured on a $RA10$ device and fits to the TA model with $H_k$ and $T$ expressed using Eq.(4) (solid line) and Eq.(5) (dashed line). The fit lines are on top of each other and indistinguishable. (b) $H_{k0}$ values obtained by fitting the data as in (a) using Eq.~(5). $RA$ dependence of (c) $J_{c0}$, (d) $\Delta$ and (d) STT efficiency obtained by fitting the data as in (a) to Eqs.(4) and (5). Each point in  (b)-(e) is the median from $\sim$30~measured cells averaged over $D$ = 60, 80, 100 and 120~nm devices.}
	\label{fig6:PW results}
\end{figure}

Fig.~6(b) shows $H_{k0}$ values as a function of $RA$. We see that, as expected, $H_{k0}$ is independent of $RA$ with $RA$-averaged values  $H_{k0}^{P \rightarrow AP} = (2.77 \pm 0.07)$~kOe and $H_{k0}^{AP \rightarrow P} = (2.32 \pm 0.12)$~kOe. These values are higher than the $H_{c0}$ values obtained from the $H$-driven magnetization reversal measurements described by Eqs.(2) and (3) (see Figs.~2 and 4), but are lower than $H_{k0}$ values obtained in thermal FMR measurements which do not involve any magnetization reversal. This is not surprising considering the different magnetization excitation and reversal processes in these measurements. Note that the difference $\left( H_{k0}^{P \rightarrow AP} - H_{k0}^{AP \rightarrow P}\right)/2  \cong 220$~Oe agrees well with the value of $H_{RL}$ obtained from fitting the data of Fig.~4.

Figs.~6(c)-6(e) compare $RA$ dependencies of $J_{c0}$, $\Delta$ and $E_b/I_{c0}$, obtained by fitting experimental data using Eqs.~(4) and (5).  We find strong $RA$ dependence of all those quantities when $t_p$ dependent $V_c$ data is fit to Eq.~(4). In particular, we observe large increase of $E_b/I_{c0}$ with increasing $RA$, similar to previous reports \cite{Hu_IEDM2017, Sun_PRB2017}.  However, when the data is fit using Eq.~(5), which takes into account VCMA and self-heating effects, all quantities become $RA$-independent. This means that STT switching parameters are intrinsically not $RA$ dependent, but their apparent $RA$ dependence is due to an error from fitting the $t_p$ vs $V_c$ assuming that STT is the only mechanism responsible for switching, without including  contributions from VCMA  and self-heating effects. 

From Fig.~6(e), the fitting model of Eq.~(5) predicts an  $RA$-independent value of $E_b/I_{c0} \cong 0.1~k_B T/\mu$A. However, from the macrospin model of Eq.~(1), taking $E_b = M_s t H_k A/2$, $E_b/I_{c0} = \hbar \eta/\left( 4 e \alpha\right) \cong 1.8~k_B T/\mu$A, using the values in Table~I. This 18 times discrepancy for $E_b/I_{c0}$ is  far greater than the aforementioned 3.5 time one for $\tau/\alpha$ despite that both expressions, derived from Eq~(1), share the same physical parameters $\hbar \eta / (2 e)$. The immediate cause of this is that the value  $\Delta \cong 70$  obtained by fitting the experimental data using Eq.~(5) (see Fig.~6(d)) is much smaller than the value $\Delta = 474$ obtained by calculating  $E_b$ using the parameter values in Table~I for average  $D = 90$~nm. Further explanations are beyond the physics of the macrospin model \cite{Sun_PRB2013, Thomas_IEDM2015}.

It is noted that the self-heating term $-\zeta V_c^2/RA'$ of Eq.~(5) explicitly violates the assumption that $E_b$ is a $T$-independent quantity, as is commonly  implied by Arrhenius-type models such as the TA model in the case of Eq.~(4) \cite{Oh_NatPhys2009}. In the Eq.~(5), the parameter  $H_{k0}$ is the room $T$ value, rather than that at $T \rightarrow 0$, and $E_b \propto H_k$  will vary with $T$ due to self-heating regardless of the presence of VCMA and STT effects. This implies that $M_s H_k$ of the cell effectively has additional $T$ dependence \cite{Thomas_IEDM2015} besides that attributable solely to thermal fluctuations in the FL magnetization direction, which is otherwise treated  by the denominator $k_B T$  in the expression for $\Delta_{eff}$ \cite{Zener_PR1954}. This could  result from the failure of the macrospin model to account for non-uniform (spin-wave mode) magnetization fluctuations.

\section{Conclusion}

In conclusion, we point out that using the obtained values for $\tau/\alpha$, $\epsilon$ and $\zeta$, we find that STT and self-heating contribute comparably to FL switching at $RA$10, and the latter is the dominant switching mechanism for larger $RA$s. As $\zeta = (R_{th}A) dH_c/dT$, higher $R_{th}A$ values should result in lower $J_c$. Two times higher $R_{th}A$ values than measured in our cells have been reported in the literature \cite{Papusoi_NJP2008, Prejbeanu_JPDAP2013}, which suggests that  further reduction of $J_c$ should be possible with thermal optimization of perpendicular MRAM cells.
		
\bibliographystyle{apsnum4-1}

\end{document}